# Dynamic Load Balancing in GPU-Based Systems – Early Experiments

Alvaro Luiz Fazenda
Institute of Science and Technology
Federal University of Sao Paulo (UNIFESP)
São José dos Campos-SP, Brazil
alvaro.fazenda@unifesp.br

Celso L. Mendes
National Center for Supercomputing Application
University of Illinois at Urbana-Champaign
Urbana-IL, USA
cmendes@illinois.edu

Laxmikant V. Kale
Department of Computer Science
University of Illinois at Urbana-Champaign
Urbana-IL, USA
kale@illinois.edu

Jairo Panetta
Computer Science Department
Aeronautics Institute of Technology (ITA)
São José dos Campos-SP, Brazil
jairo.panetta@gmail.com

Eduardo Rocha Rodrigues
IBM Research Brazil
Rio de Janeiro-RJ, Brazil
edrodri@br.ibm.com

**The dynamic load-balancing framework in Charm++/AMPI, developed at the University of Illinois, is based on using processor virtualization to allow thread migration across processors. This framework has been successfully applied to many scientific applications in the past, such as BRAMS, NAMD, ChaNGa, and others. Most of these applications use only CPUs to perform their operations. However, the use of GPUs to improve computational performance is quickly getting massively disseminated in the high-performance computing community. This paper aims to investigate how the same Charm++/AMPI framework can be extended to balance load in a synthetic application inspired by the BRAMS numerical forecast model, running mostly on GPUs rather than on CPUs. Many major questions involving the use of GPUs with AMPI where handled in this work, including: how to measure the GPU's load, how to use and share GPUs among user-level threads, and what results are obtained when applying the mandatory over-decomposition technique to a GPU-accelerated program.**

## I. INTRODUCTION

According to Jack Dongarra [1], load imbalance is one of the major problems to be handled to improve the parallel performance of applications. The other main reasons are serial phases of the code (related to Amdhal law) and communication overhead. The load imbalance problem has been studied for a long time, and many different strategies to solve it were created along the years.

Many of the techniques to balance the overall computational loads in parallel runs are embedded into applications, but, in the context of this work, a framework scheme will be analyzed and tested, allowing to isolate the application from the balancing strategies. The Charm++/AMPI is a framework for dynamic load balancing that uses process virtualization with thread migration. This approach is convenient to large applications because it demands minimal changes in the legacy code.

The idea of process migration consists in a run-time system able to migrate threads from most loaded processors to less loaded ones, every time a load imbalance reaches some defined threshold, measured at user' defined places in the source code. In this approach, each processor is in charge of a large number of threads, which could migrate if necessary.

The NAMD Molecular Dynamics model [2, 3] is the best example for success of the Charm++ framework. The recent 2012 IEEE Computer Society Sidney Fernbach Award, "for outstanding contributions to the development of widely used parallel software for large biomolecular systems simulation" is a proof of that. In that application, the Charm++ framework allows the NAMD model to scale to obtain results allowing to describe the complete atomic HIV-1 capsid (protein shell of a virus) model [4]. Recently, the same research group at University of Illinois obtained results running the NAMD model in a hybrid system composed of CPUs and GPUs [5], however the dynamic load balancing is applied only to the procedures running on CPUs, with the loads remaining fixed in the processes on the GPUs.

Kunzman [6] obtained good results extending the Charm++ balancers and run-time system to be able to deal with hardware accelerators like Cells and GPUs (using CUDA programming). The strategy includes two-phase balancing procedures: the first phase to balance the load between CPU and GPU, and the second phase to balance threads' loads among all processors.

When using Charm++, the original program usually requires few modifications to be designed in a message-driven execution model. For huge and complex source-codes designed with message passing scheme in usual MPI, the Adaptive MPI, or AMPI [7], seems to be the obvious choice. According to Zheng et all [8], AMPI is an adaptive implementation and extension of MPI built on top of the Charm++ run-time system. AMPI implements virtualized MPI processes (VPs) using light-weight migratable user-level threads, several of which may be mapped to one physical processor. AMPI nowadays could be considered a mature implementation of MPI; it has been used in real world applications, such the FEM framework [9] in a dynamic 3D crack propagation simulation program, and with BRAMS (Brazilian developments on the Regional Atmospheric Modeling System) [10].

All these applications that received benefits from AMPI did not use any hardware accelerator to improve numerical performance, despite their massively disseminated use in the high-performance computing community nowadays, especially with GP-GPUs (General-purpose computing on graphics processing units) programming. In hybrid system programming, it is worth to note the possible adoption of the OpenAcc application program interface, which describes a collection of compiler directives to specify regions of code written in standard C/C++ and Fortran to be offloaded from a host CPU to an attached accelerator, in a similar way to the well know OpenMP directives. This new programming standard could improve the porting of source-codes to use hardware accelerators, making the numerical power of GPUs easy to use in large codes.

The motivation to this work is to discover how could be possible to use the AMPI framework, and the domain overdecomposition whith thread virtualization/migration, to perform the dynamic load balance in a synthetic scientific application inspired by the BRAMS meteorological model, fully accelerated by GPUs through OpenAcc's API codification.

This paper is organized in the following manner: the next section briefly shows some related work using different techniques to dynamic balance systems running in accelerators, the third section describes the Charm++/AMPI framework; the forth section shows the background used in the numerical simulation. The fifth section discusses the GPU performance and the reliable way to make the load measurements. The computational results can be found in the sixth section and the conclusion in the last section.

## II. Related Work

Others related works were developed using different methods to balance programs running in accelerators. Some examples of that could includes the work of Chen et all [17], who developed a task-based dynamic load-balancing solution for single-and multi-GPU systems that make possible to use the device hardware more efficiently than the CUDA scheduler for unbalanced workload in their benchmarks. Robert Hagan in his Master thesis [18] developed methods for data-driven and dynamic multi-GPU load balancing using a pipelined approach and a framework for use with different applications. The dynamic load balancing method developed is based on buffer fullness and can adjust to workload changes at runtime to gain an additional performance improvement. Costa et al [19] developed a dynamic load balancing library that allows parallel code to be adapted to a wide variety of heterogeneous systems with minimal overhead and negligible cost to the programmer. The case studied by them include matrix multiply and resource allocation problems in different heterogeneous scenarios in multi-GPU systems.

It is also worth to notice the work made by J. Dongarra group in the development of MAGMA [20], an extension of the well known PLASMA library used as a solver for dense numerical linear algebra problems in multi-core systems, which balance the process load under accelerators by using the well established DAG computational model with a new dynamic task scheduling.

## III. Dynamic Load Balancing by Process Virtualization/Migration

Charm++ is an object-oriented parallel programming system, developed at the University of Illinois in Urbana-Champaign, aimed to improve the productivity in parallel programming while enhancing scalable parallel performance [11]. Charm++ employs the idea of overdecomposition with processor virtualization based on migratable objects. In this approach, the programmer decomposes the application's data into a number of objects, called VPs (Virtual Processes), which is larger than the number of available processors, allowing a specific program to express much more parallelism than the available number of processors in a parallel context, thereby creating an abundance of parallelism that can be exploited.

Those VPs are automatically mapped to the available processors by the Charm++ runtime system, and they can migrate across processors during execution to balance the overall load. Each of the VPs residing on a certain processor is implemented as an user-level thread, which ensures fast context-switch, and is given control of the CPU by a local scheduler provided by Charm++, in a non-preemptive fashion [12]. This scheme greatly improves the overlap between computation and communication.

Adaptive MPI (AMPI) is a full MPI implementation built on top of Charm++ [13]. In AMPI, each MPI "rank" is implemented as a user-level thread embedded in a Charm++ virtual processor (VP). This approach ensures that the benefits of Charm++ are available to the broad class of applications written using MPI. Thus, an MPI program designed to be run on K processors is typically executed by AMPI with K virtual processors on P physical processors, where usually K>P.

A potential problem that could be faced when using VPs in an AMPI program is related to possible conflicts involving static and global variables, since the VPs are implemented as threads that share the same address space, in contrast with MPI programs where each process has its own private address space. There are a few different mechanisms for variable privatization, with varying degrees of automation and overhead [14]. A simple and usual solution consists in using a flag (swapglobals, applied to platforms that support ELF format) that ensure a private global variable space to each thread but, unfortunately, this method does not work for static variables, which should be handled by the programmer, probably by transforming those statics in global variables. Other methods include a compiler-based refactoring technique, which could use some tool to automate the procedure, like Photran, an Eclipse plugin allowing handling Fortran codes, and an automatic technique, similar to "swapglobals" flag, based on thread local storage (TLS), but restricted until now to a specific compiler using static linking, which unfortunately prevents GPU processing.

IV. BACKGROUND: USING OPENACC IN GPGPU COMPUTATION FOR A SYNTHETIC SCIENTIFIC COMPUTING APPLICATION

To perform the computational tests, a synthetic application was developed based on the meteorological model BRAMS. Some of the major characteristics present in that meteorological model, and also present in most of other models, were simulated by simpler codes. Therefore, the synthetic application simulates the most numerical intensive phases responsible to solve the fluid dynamics and cloud physics by a 3D Jacobi solver, which is pretty similar to the laminar diffusion procedure in fluid dynamics, and by a synthetic code including flux dependency in the vertical axis, similar to Cumming et al [15] to represent the same phenomena in COSMO meteorological model, and illustrated in Figure 1 respectively.

```
do j=1,Ny
   do i=1,Nx
      do k=2,Nz
         a(k,i,j)=f(b(k,i,j), a(k-1,i,j))
      end do
   end do
end do
```

Fig. 1. Synthetic code to simulate cloud physic

The 3D Jacobi solver implies in creation of a lateral boundary exchange procedure implemented in a 2D domain decomposition, like in BRAMS. An artificial load imbalance was forced in the physic through a similar procedure, trying to simulate what usually occurs in the BRAMS model when precipitation is present [12]. The numerical procedures (3D Jacobi and Physics) will run in the GPU accelerator by using OpenAcc directives applied to source-code. The data transfers between CPU and GPU were minimized for obvious performance reasons, however CPU-GPU transfers happen in each timestep for the data related to the boundary of sub-domains in the MPI message passing procedure, since GPU to GPU direct data transfer was not used yet (like the CUDA-Aware MPI). A full data transfer between CPU and GPU happens only when the migration is needed, which occurred in a couple of timesteps, since there are no mechanisms to migrate data held on acceleration devices by Charm++/AMPI. Figure 2 presents a source-code to summarize the algorithm for the depicted synthetic application that we developed.

The computational infra-structure used in the tests consists in a Cray XK7 machine with 8 nodes linked by Gemini network, each one configured with an Interlagos 16-core AMD Opteron CPU, 17 Gbytes of main memory and 1 Kepler K20 NVIDIA Tesla GPU device. All programs were compiled using Cray Fortran and C compilers, and the latest Charm++ version.

The MPI interface was generated by AMPI, which uses Charm++ primitives.

```
Initialization
do mig=1,maxmig !migration loop
   Transfer full data to GPU (CPU->GPU)
   do timestep=1,stepsbetmig
      If (measurement_timestep)
         INSTRUMENT(ON)
         mode = 0 !syncronous mode
      Else
         INSTRUMENT(OFF)
         mode = 1 ! Assyncronous mode
      Endif
      compute boundaries
      update boundary on host (GPU->CPU)
      Comm. boundaries by MPI non-block
      Launch JACOBI kernel
      if (mode==1)
         context change for user-threads
      Launch PHYSIC kernel
      if (mode==1)
         context change for user-threads
      MPI_WAITs
      update boundary on device(CPU->GPU)
   Enddo
   Transfer full data to CPU (GPU->CPU)
   MPI_MIGRATE
Enddo
```

Fig. 2. Synthetic application Algorithm

V. GPU PERFORMANCE AND LOAD MEASUREMENTS

The Charm++/AMPI framework is based on load measurements to make the necessary migrations to keep the parallel program balanced. Until the Charm++ version 6.5.0, only CPU loads are considered by the balancing algorithm. Thus, a reliable way to measure the load on the accelerated device must be created. An obvious choice is to run jobs on GPUs in a synchronous mode, where the host is just waiting the completion of procedures on the accelerator. In this case, the CPU wall time must be quite similar to GPU elapsed time. Other ways to measure the GPU load in real-time consists in using CUDA events or PAPI hardware counters. All those techniques show similar and reliable results when synchronous kernel launches are used.

However, since there are several threads (user-threads) running in a process for an AMPI/Charm++ program, multiple streams could be processed by the GPU concurrently, considering that this feature is supported by modern NVIDIA GPU devices like Tesla, Fermi and Kepler models. This characteristic could improve the overall performance, allowing a better occupancy of GPUs, many times by overlapping data transfers with processing. To run multiple GPU streams each user-thread (or VP) could simply launch an asynchronous GPU kernel and invoke a context switch. This context switch could be made by calling a "local barrier", i.e. perform a barrier only in the scope of user-threads for a specific process. In that case, the Charm++ runtime system will force the context to be switched to the next user-thread that still did not reach the local barrier. The function AMPI_YELD, available in the AMPI library, implements the required context-switch.

Another possibility to explore concurrent GPU kernels consists in sharing the GPU among multiple processes in an MPI code, which is available in the NVIDIA Kepler devices that implement HYPER-Q [16] and the CUDA-Proxy feature. However, concurrent GPU kernels (sharing resources among user-threads or regular processes) show an unreliable load measure. In an asynchronous kernel launch, the CPU wall time cannot be used, because it just returns the elapsed time spent to launch the kernel on device, instead of the required time for completion of the kernel. Cuda events and PAPI counters return information not considering the different Kernel contexts running on the device, which implies in spurious elapsed times related to each job. It is worth to notice that PAPI counters do not work with multiple processes sharing the GPU by CUDA-proxy, but can be used with multiple threads.

Considering all these characteristics related to GPU load measurement, the only reliable way to collect the required load consists in using the synchronous kernel launch. The method to measure could be either CPU wall time, CUDA event or PAPI hardware counter, since there is not any concurrent job running on the GPU. Launching concurrent asynchronous GPU kernels could improve the overall performance in a process with multiple user-threads, but imposes an unreliable load measurement. The strategy adopted in the application developed consists in running as much as possible with asynchronous concurrent jobs and switch to a synchronous serial job launch in determined timesteps, when measurements are taking place. Table I shows wall times (average elapsed time per timestep) to run the application in a full domain with 1024x1024x40 array size and 100 different variables representing the physics fields present in a meteorological numerical simulation, with and without concurrent asynchronous GPU jobs.

The first two rows on Table I show results considering 2 processes (P=2) total, one per cluster node, and each process with 2 user-threads (VP=4) sharing the same GPU. The third and fourth rows in Table I show results for regular MPI executions and should be used just for comparative purposes, since those configurations do not allow any thread migration. The third row shows the result with 4 processes (P=4) total in regular MPI mode, using Hyper-Q and the Cuda-Proxy feature to allow multi processes sharing one GPU, showing the best performance for the same configuration and domain decompostion within the same computer resources, without the overhead for thread switches. The last row in the same Table I shows the result if using just 2 processes at all, one of them per node, representing the normal case where there are no extra threads per process.

The mode with multiple processes sharing a GPU (by using HYPER-Q and Cuda-Proxy) was not considered in this early experiment due to problems to obtain a reliable GPU load measurement. One possible option to solve this problem consists in implementing a critical section to force serialization among the multiple processes in the same node, which could be activated just at the periods chosen for the measurements.

TABLE I. RESULTS IN SYNCHRONOUS AND ASYNCHRONOUS MODE

| Mode | Average Time (s) per timestep |
|---|---|
| Synchronous | 12.3 |
| Asynchronous | 11.6 |
| P=VP=4 (2 processes per node) | 11.0 |
| P=2 (1 process per node) | 11.8 |

The Charm++ balancing strategies imply in using an domain over-decomposition, as already depicted before. It is worth to notice that the over-decomposition technique works as expected only if the scalability with problem size is approximately linear. This could be better understood imagining that when any problem is decomposed by two, creating new problems with approximately half the orginal problem size, the two new programs should run in half of the time, approximately.

The parallel program scalability with problem size depends on many issues, but the algorithm complexity and communication overhead could be considered the main factors. However, when using GPUs to perform parallel processing a different behavior was found for some specific algorithms over-decomposed, when comparing to the same algorithm running in a regular CPU. Figure 3 shows an example of source-code where the scalability with problem size is different for CPU and GPU. This source-code includes OpenAcc directives to run in parallel on the GPU. Focusing only in the parallel processing on device and not considering the data transfer, only the first two loops are distributed by the several parallel cores on the GPU, while the inner loop is performed serially by each GPU core.

```
!$acc kernels
!$acc loop collapse(2)
do j=2,m-1
   do i=2,n-1
     !$acc loop seq
     do k=1,innerloopmax
        B(i,j) = B(i,j) + &
                0.25*(a(i-1,j) + &
                a(i+1,j) + a(i,j-1) + &
                a(i,j+1))
     enddo
   enddo
enddo
!$acc end kernels
```

Fig. 3. Source-code example with OpenAcc directives for testing scalability with problem size

The results for CPU and GPU for the example algorithm are shown in Table II. In this table, it is possible to confirm the linear scalability with problem size when running on CPU (fourth column), however the same program running in parallel on GPU shows an unexpected result (fifth column), which could decrease the optimization gain obtained by dynamic load balance with thread migrations. This undesired behavior on GPU is due to the inner loop being executed in serial mode, since there is a fixed time to process it, even if using multiple GPU cores. Fortunately, this problem did not happen in a significant manner in the source-code of the application developed, but it shows that users must be on alert for that scaling pattern with problem size on their codes, especially when running on a GPU device.

TABLE II. SCALABILITY WITH PROBLEM SIZE - RESULTS ON CPU AND ON GPU

| N | M | Inner | CPU time (s) | GPU time (s) |
|---|---|---|---|---|
| 1024 | 512 | 200000 | 54.41 | 0.82 |
| 1024 | 256 | 200000 | 27.1 | 0.49 (59,5%) |
| 1024 | 128 | 200000 | 13.45 | 0.33 (67%) |
| 1024 | 64 | 200000 | | 0.17 |
| 1024 | 32 | 200000 | | 0.18 |

## VI. EXPERIMENTAL RESULTS

Three experiments was performed in order to validate the strategy developed allowing dynamic load balancing with thread virtualization/migration in a GPU accelerated application:

A) Static imbalance with 4 sub-domains (and 4 VPs);

B) Dynamic imbalance (but reproducible) with 8 sub-domains (and 8 VPs), using two different balancer algorithms;

C) Dynamic imbalance (but reproducible) with 16 sub-domains (and 8 VPs), using two different balancer algorithms.

For the first experiment (A) the same configuration was used for the synthetic application depicted in section IV: a 1024x1024x40 problem size and 100 different variables representing the physics fields. The strategy to maximize the device performance through concurrent asynchronous kernel launch was implemented in the following manner: between calls to function MPI_Migrate (source-code in Figure 2, which is responsible to activate the balancer), the program performs several (N) timesteps launching concurrent GPU kernels in asynchronous mode among the user-threads, and performs some (M) timesteps in synchronous mode to obtain a reliable load measurement. These synchronous kernel launches are activated in the final timesteps. Thus, for the experiment A, the first 15 timesteps were performed in an asynchronous way and the last 5 timesteps in a synchronous way to reliably measure the load. However, the time interval, quantity of synchronous steps and their frequency should be determined according to the problem needs.

An artificial load imbalance of approximately 50% was forced on the physics phase (Figure 1) between the most and the least demanding numerical procedure, by changing the loop trip on the inner loop on specific threads. For this first experiment, the process and their aggregated threads on node zero received the most demanding load. Running with only 2 MPI regular processes, one per node, results in an execution time of 236.5 seconds for the 20 timesteps performed. This execution time could be used to compare the results with the case where it is possible to balance load using thread migration. Using the AMPI code with 4 sub-domains (2 processes at each node, each one with 2 VPs) launching GPU jobs in an asynchronous way for the first 15 timesteps and in a synchronous way for the next 5 timesteps, the execution time was 231.4 seconds. After this first 20 timesteps, the Charm++ runtime, through a specific balancing algorithm called "GreedyLB" [12] which always assigns the heaviest object to the least loaded processor, migrated two threads to balance the overall load (one thread from node 0 moves to node 1 and vice-versa) and the resultant execution time was 168.9 seconds for the next 20 timesteps performed. All this information related to performance and other measurements can be checked on Table III.

The results for the second and third rows on Table III show that the dynamic load balancer is working as expected for an AMPI code running on GPUs. The performance obtained by the strategy of asynchronous kernel launches on GPU among the user-threads showed a results close to the best performance obtained when using multiple processes sharing GPUs in a normal MPI code with a load distribution conveniently forced to be balanced. In fact, the solution developed with AMPI is just 3.6% worse compared to this forced balanced case, which validates the technique developed.

TABLE III. RESULTS FOR THE DYNAMIC LOAD BALANCER – EXPERIMENT A

| Mode | Time (s) |
|---|---|
| P=2 (1 process per node) unbalanced | 236.5 |
| First measurement unbalanced | 231.4 |
| Second measurement after balancing by GreedyLB | 168.9 |

The next experiment (B) consists in simulating a problem with a dynamic imbalance, instead of the static imbalance of the previous experiment. To achieve this, an artificial array controlling the load on the physics code was implemented to define the inner loop trip in each horizontal position on the original code of Figure 1. This inner loop was implemented as a serial loop due to a flux dependency on the first index, representing the vertical axis in a 3D atmosphere, like found in the BRAMS model. The resultant source-code is presented on Figure 4.

```
!$acc kernels present(A,B,C)
!$acc loop collapse(2) private(i,j,k,kr)
do j=1,myp
   do i=1,mxp
      !$acc loop seq
      do k=2,mzp*C(1,i,j)
         if (k>mzp) then
            kr = k - (int(k/mzp)*mzp)
         else
            kr = k
         endif
         A(kr,i,j) = f(B(kr,i,j),
                       A(kr-1,i,j))
      enddo
   enddo
enddo
!$acc end kernels
```

Fig. 4. Synthetic code to simulate cloud physics with imbalance controlled by an array

The initial values for each element on array C (which control the inner loop for the code on Figure 4) were set to one for the least demanding thread and to two for the most demanding thread, depending on where the thread was located in the 2D decomposition. A specific procedure was also created to move (similar to advection transport in fluid dynamics) the value in the array C through the entire domain among threads or sub-domains. In early experiments, a 1D domain decomposition was imposed to the problem, distributing sub-domains over the Y axis, like in Figure 5.

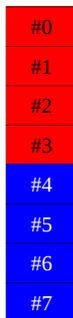

Fig. 5. 1D initial domain decomposition

Considering the domain decomposition like in Figure 5, where 8 sub-domains were created and equally distributed as user-threads in 4 processes, the first upper half (sub-domains #0 to #3) of the entire domain received initially the most demanding load (represented by the red color), and the lower half received a least demanding load (blue color). Over the timesteps, these initial load distributions will move in a upper-down direction, finishing with the load distribution on Figure 6, where the red and blue sub-domains represent the most and least demanding load, respectively.

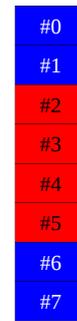

Fig. 6. 1D final domain decomposition

In this experiment (B), just 50 3D arrays are used as generic meteorological fields, and the application ran 40 timesteps total. The interval between possible migration is just 10 timesteps, where the first 6 timesteps were performed in asynchronous mode and the last 4 timesteps in synchronous mode. After each 10 timesteps a call to the balancer through MPI_Migrate is performed, which could imply in thread migrations.

The first 10 timesteps occurred when the system was unbalanced in the initial state (Figure 5). The next timesteps (between 11 and 20) were performed after a possible thread redistribution. After that point, another call to the balancer is performed, followed by a new load configuration equal to the final state showed by Figure 6. Ten more timesteps (between 21 and 30) are performed with this new load configuration, and a new call to the balancer is performed, followed by the final 10 timesteps.

For the first call to the balancer, the "GreedyLB" balancer was used. For the second call, another balancer was called, named "RefineSwapLB" which moves objects away from the most overloaded processors to reach average. This balancer algorithm can also swaps objects to reduce the load on the overloaded processor, in case it cannot migrate an object from an overloaded processor to an underloaded processor. Both calls reached the expected result, invoking thread migrations to keep the system balanced. The execution time for this second experiment can be checked on Table IV, where the second column shows the executions times for 10 timesteps.

TABLE IV. RESULTS FOR THE DYNAMIC LOAD BALANCER - EXPERIMENT B (P=4, VP=8)

| Timestep Interval | Time (s) measured (10 timesteps) |
|---|---|
| 1-10 | 28.36 |
| 11-20 (after first migration) | 23.10 |
| 21-30 (after load redistribution) | 28.10 |
| 31-40 (after second migration) | 23.00 |

Extending this previous experiment to the experiment c, the same program configuration was used with more VPs per process: using 4 processes, each one with 4 user-threads, resulting in 16 sub-domains. The artificial dynamic imbalance follows the same pattern described in the second experiment. The results for this third experiment (C) are on Table V, where it is possible the execution times for each 10 timestep interval and the thread distribution along the same intervals. In the thread distribution each thread number identifies the node where it was originally located and the color identifies the load: red for heaviest threads and blue for light ones. Each set of threads separated by an space is running on a node.

TABLE V.  RESULTS FOR THE DYNAMIC LOAD BALANCER - EXPERIMENT C (P=4, VP=16)

| Timestep Interval | Time (s) measured (10 timesteps) | Thread distribution |
|---|---|---|
| 1-10 | 27.10 | 0000 1111 2222 3333 |
| 11-20 (after 1st migration) | 23.00 | 2201 0313 1231 0203 (12 threads migrated) |
| 21-30 (after load redistribution) | 24.78 | 2201 0313 1231 0203 |
| 31-40 (after 2nd migration) | 22.50 | 2200 1313 1230 1203 (4 threads migrated) |

In the experiment C (Table V) the first call to MPI_Migrate, just after the timestep 10, caused 12 thread migrations in order to balance the overall load through the "GreedyLB" balancer. After the new load distribution in the end of timestep 20, this balanced application became imbalanced again, since the first and third processes hold 3 heavy threads and just 1 light thread, and the second and forth processes hold the opposite configuration (1 heavy and 3 light VPs). The execution time for this new load pattern is greater, as expected, than in the previous interval. However, the Charm++ runtime system balanced the load again, promoting 4 thread migrations by "RefineSwapLB" balancer, which optimized the execution time to the same level as found for timesteps interval between 11 and 20.

## VII. CONCLUSIONS

The early experiments validate the strategy developed to make possible to use the Charm++/AMPI framework on an application fully accelerated by GPUs, showing expected thread migrations in the presence of dynamic load imbalance. The strategy consists in launch concurrent kernels on device to improve performance and launch serial kernels only when and where is necessary to make reliable GPU load measurements. The results in Table IV and V assure that the method developed was able to balance the system over different load and sub-domains configurations by thread migrations. Comparing experiments B and C it is possible to notice no significantly execution time difference when using more VPs for the same problem configuration.

Two different balancers were successfully tested in the experiments. In the Table V, the " GreedyLB" balancer was used for the first migration, and it addressed 12 VPs migrations to balance the system, distributing 2 heavy and 2 light VPs per process. In fact, only 8 VPs migrations are needed to balance the overall load, but the "GreedyLB" balancer migrated more VPs than the necessary, which should imply in some overhead, however, other tests with different balancers did not reach better results than that, and this balancer was used just for the first time.

The "RefineSwapLB" balancer, when compared to "GreedyLB", shows a more conservative thread migration. If it was used for the first balancing round, in the experiment C, it migrated less VPs than the necessary to keep the system balanced. Thus, for these experiments the best option were to use a more aggressive migration algorithm like "GreedyLB" for the first call to MPI_Migrate function and a more conservative balancer, like "RefineSwapLB", for the subsequent calls to the same function. With this scheme (GreedyLB + RefineSwapLB) it was possible to imposes an aggressive migration procedure in the first round when the system is totally imbalanced, and uses a more conservative migration algorithm in the next timesteps, when is important to avoid unnecessary migrations made by "GreedyLB".

A plan for future development, extending this paper, consists in implementing an algorithm to minimize the data transfers between CPU and GPU, detecting this necessity only when the imbalance measured among processes (not threads) is beyond a threshold. In the case of necessity, a next timestep with a reliable GPU job launch will be performed to detect what specific threads should be migrated. Other future developments also include porting parts of the BRAMS code and testing the same strategy under a real dynamic load problem, and creating a method that uses over-decomposition to distribute problems between CPU and GPU, instead of using only GPU jobs like in these early experiments.

ACKNOWLEDGMENT

This work was supported by FAPESP – Sao Paulo Research Foundation.